\documentclass[runningheads,a4paper]{article}
\usepackage{graphicx,graphics,color}
\usepackage{natbib}
\usepackage{array}
\usepackage{booktabs}
\usepackage{tabularx}
\usepackage{amsmath}
\usepackage{etoolbox}

\usepackage{fancyhdr}
\usepackage{breqn}

\providecommand\bnabla{\boldsymbol{\nabla}}

\newsavebox{\astrutbox}
\sbox{\astrutbox}{\rule[-5pt]{0pt}{20pt}}

\newcommand\md{\mathrm{d}}

\def\nablab{{\mbox{\boldmath $\nabla$}}}

\usepackage{color}

\definecolor{darkgreen}{rgb}{0,0.4,0}
\definecolor{titlecolor}{rgb}{0.9,0,0}

\title{Upper bound of heat flux in an anelastic model for Rayleigh-B\'enard convection}
\author{T. Alboussi\`ere, Y. Ricard and S. Labrosse \\[3mm] {\small \sc Laboratoire de G\'eologie de Lyon}}
\date{\small \today}

\begin{document}

\maketitle

\begin{abstract}
	Bounds on heat transfer have been the subject of previous studies concerning convection in the Boussinesq approximation: in the Rayleigh-B\'enard configuration, the first result obtained by \cite{howard63} states that $Nu < (3/64 \ Ra)^{1/2}$ for large values of the Rayleigh number $Ra$, independently of the Prandtl number $Pr$. This is still the best known upper bound, only with the prefactor improved to $Nu < 1/6 \ Ra^{1/2}$ by \cite{DoeringConstantin96}. In the present paper, this result is extended to compressible convection. An upper bound is obtained for the anelastic liquid approximation, which is similar to the anelastic model used in astrophysics based on a turbulent diffusivity for entropy. The anelastic bound is still scaling as $Ra^{1/2}$,  independently of $Pr$, but depends on the dissipation number $\mathcal{D}$ and on the equation of state. For monatomic gases and large Rayleigh numbers, the bound is $Nu < 146\, Ra^{\frac{1}{2}} / (2-\mathcal{D} )^{\frac{5}{2}}$. 
\end{abstract}

\section{Introduction}

An important landmark of fluid mechanics has been to show that rigorous upper bounds could be obtained from the governing equations on quantities such as energy dissipation (or pressure gradient) in a pipe with a given flow rate, or heat flux between walls maintained at different temperatures \citep{howard63}. Concerning Rayleigh-B\'enard convection, 
the method of \cite{howard63} consisted in identifying an integral equation based on energy conservation restricting the space of possible temperature fields, and then finding an upper bound on the heat flux among fields in that restricted space. A second method -- the so-called 'background' method -- was developed in the 90s by \cite{DoeringConstantin96} and is based on a decomposition of the temperature field into an arbitrary vertical profile (the background profile) satisfying the boundary condition and an homogeneous 3D, time-dependent field. A spectral condition is said to hold when the 'dissipation' contained in the background profile (in fact the $L^2$ norm of its derivative) is larger than the total possible dissipation of the convective flow. This spectral condition has been shown to be related to the same eigenvalue problem as that involved in the energy stability \citep{joseph76} of the background profile. The problem is finally turned into finding the background profile with the minimum possible dissipation. A third method was obtained recently by \cite{seis2015}, with a more intuitive approach. The average heat flux must be constant over height, but cannot be carried by conduction after some distance to the bottom wall and convection must take over. This implies that sufficiently strong vertical velocity components exist there, which is necessarily associated with deformation (hence viscous dissipation) as those vertical components are zero at the bottom. However the total viscous dissipation is related to the heat flux and imposes a limit to the convective flux. That constraint leads to the same scaling as that of Howard.

Upper bounds of the heat flux have not been derived for compressible convection until now. Here, in section \ref{goveq}, we consider a simple model of compressible convection, the anelastic liquid approximation \citep{ajs2005}. As an anelastic model, acoustic modes have been filtered out. Moreover, entropy is supposed to depend on the superadiabatic temperature only, so that pressure is not a relevant thermodynamic variable. In astrophysics, the anelastic model is used too \citep{lf1999} with nearly the same equations as the anelastic liquid model, although the path to get there has taken a different direction. From a general anelastic model, a subgrid model for turbulence is used to change the conduction term (gradient of temperature) into a gradient of entropy. Again, this anelastic model depends only on a single thermodynamic variable, entropy. We also use a simple equation of state, that of the ideal gases. Finally, we consider a simple geometry, that of a plane layer, in a uniform gravity field perpendicular to the plane layer. The horizontal extent of the layer can be infinite or finite. The vertical depth of the layer is such that compressible effects range from negligible (Boussinesq limit) to extreme values (adiabatic temperature profile reaching zero kelvin at the top). 

The maximum principle for parabolic equations plays an important role in our derivation of an upper bound. This also plays a crucial role in the work by \cite{seis2015}. In the Boussinesq model, temperature is bounded below by the cold temperature imposed at the top and bounded above by the hot temperature imposed at the bottom. In a compressible model, adiabatic compression and decompression as well as viscous heating imply that these limits not longer hold for temperature. Instead, we show in section \ref{minpple} that entropy has a minimum value imposed at the top boundary but no obvious maximum value. That property will be used several times in the paper. 

In section \ref{seclogentropy} we derive an equation for the logarithm of entropy (up to a constant), a quantity that we call $\log$-entropy. We show that, similarly to the entropy flux, the flux of that $\log$-entropy increases with height. Otherwise stated, the sources of $\log$-entropy are positive. With that equation, we can bound the gradients of the $\log$-entropy in the layer. The derivation follows then the same principle as that of \cite{seis2015} and a lower bound of kinetic energy is found to be necessary to carry the  flux of $\log$-entropy. Coming back to the entropy equation (not $\log$-entropy) in section \ref{viscousdiss}, we obtain an upper bound for dissipation. As shown in section \ref{upperbound}, the condition that the lower bound is less than the upper bound of dissipation leads to an upper bound for the heat flux in terms of the governing parameters.  


\section{Governing equations}
\label{goveq}

The fluid (a monatomic ideal gas) is contained in a horizontal layer, between altitudes $0$ and $d$, in a uniform gravity field ${\bf g}$. A superadiabatic temperature difference $\Delta T_{sa}$ is imposed in addition to the adiabatic gradient between the bottom and top boundaries. 
The governing equations in the anelastic liquid approximation are written in a dimensionless form as follows \citep[see][]{ajs2005}
\begin{align}
	& \nablab \cdot (\rho_a \mathbf{v})  = 0	, \label{diva}	\\
	& \frac{\rho _a}{Pr} \frac{\mathrm{D} \mathbf{v}}{\mathrm{D} t} = 	- \rho_a \nablab\left( \frac{ P}{\rho_a}\right) + Ra \rho_a s \mathbf{e}_z + \nablab \cdot \tau , \label{momentuma} \\
	& \rho_a T_a \frac{\mathrm{D}  s}{\mathrm{D} t} = \frac{\mathcal{D}}{Ra } \dot{\varepsilon} : \tau + \nabla^2 T , \label{entropya}
\end{align}
where $\mathbf{e}_z$ is the vertical unit vector ($\mathbf{e}_x$ and $\mathbf{e}_y$ are the horizontal unit vectors) and where the dimensionless governing parameters are the Prandtl $Pr$, Rayleigh $Ra$ and dissipation $\mathcal{D}$ numbers
\begin{align}
       & Pr = \frac{\eta c_p}{k}, \hspace*{5mm} Ra= \frac{\rho _0^2 c_p g \Delta T _{sa} d^3}{T_0 \eta k}, \hspace*{5mm} \mathcal{D} = \frac{g d}{c_p T_0}, \label{adimPar}
\end{align}
where the viscosity $\eta$, the thermal conductivity $k$ and the heat capacity $c_p$ of the gas are uniform and constant. The dimensionless tensors of deformation rate and stress, in the Stokes approximation of zero bulk viscosity \citep[proven correct for monatomic ideal gases][]{EMANUEL19981313}, are the following
\begin{align}
       &\dot{\varepsilon} _{ij} = \frac{1}{2} \left( \partial _i v_j + \partial _j v_i   \right),  \label{def} \\
       &\tau _{ij} = 2 \dot{\varepsilon} _{ij} -  \frac{2}{3} ( \partial _k v_k) \delta _{ij}. \label{stress}
\end{align}
The average temperature $T_0$ and average density $\rho _0$ of the adiabatic profiles are chosen to express dimensionless temperature and density adiabatic profiles (hydrostatic, isentropic) as follows \citep{cdadlr2019}
\begin{align}
 T_a (z) &= 1 - \mathcal{D} \left(z - \frac{1}{2} \right) , \label{Ta} \\
	\rho _a (z) &= \frac{\mathcal{D}/ \left( 1 - \gamma ^{-1}  \right) }{\left( 1+\mathcal{D}/2  \right) ^{\frac{\gamma}{\gamma -1}} - \left( 1-\mathcal{D}/2  \right) ^{\frac{\gamma}{\gamma -1}}} \left[ T_a (z) \right]  ^{\frac{1}{\gamma -1}}, \label{rhoa} 
\end{align}
where $\gamma = c_p/c_v$ is the ratio of heat capacities (for example $\gamma = 5/3$ for monatomic gases). The dimensionless gradient of adiabatic temperature is $- \mathcal{D} {\bf e}_z$, uniform and vertical. 

Superadiabatic temperature $T$ and entropy $s$ are scaled using $\Delta T _{sa}$ and $c_p \Delta T_{sa}/T_0$, respectively. Space coordinates $(x,y,z)$, time $t$, velocity $\mathbf{v}$, rate of deformation tensor $\dot{\varepsilon}$, stress tensor $\tau$ and pressure $P$ are made dimensionless using $d$, $\rho _0 c_p d^2 / k$, $k/(\rho _0 c_p d)$, $k/(\rho _0 c_p d^2)$, $k \eta /(\rho _0 c_p d^2)$ and $k \eta /(\rho _0 c_p d^2)$ respectively. 

In the anelastic liquid model, entropy $s$ -- or rather the superadiabatic entropy in addition to a uniform base value -- is assumed to depend on superadiabatic temperature $T$ only
\begin{equation}
	s = \frac{T}{T_a} .\label{Ts}
\end{equation}
A consequence is that pressure $P$ has no effect on thermodynamic variables. In this model, pressure $P$ is only a Lagrange multiplier associated with the conservation of mass (\ref{diva}).

The boundary conditions are given by constant values of temperature or entropy on bottom and top boundaries. In terms of velocity, we impose that the normal component vanishes on both horizontal boundaries and that no work is done by the boundaries. Both non-slip (zero tangential velocity components) or no-stress in the horizontal direction are acceptable.
\begin{align}
	z&=0: \hspace*{3 mm}	v_z = 0 \hspace*{3 mm} \mathrm{and}  \hspace*{3 mm}( v_x = v_y = 0  \hspace*{3 mm}\mathrm{or}\hspace*{3 mm} \partial _z v_x = \partial _z v_y = 0),   \label{bottomTsv} \\
	z&=0: \hspace*{3 mm}  T = \frac{1}{2} \hspace*{3 mm}\mathrm{or}\hspace*{3 mm} s = \frac{1}{2 + \mathcal{D}}, \hspace*{3 mm} \mathrm{since}\hspace*{3 mm} T_a = 1 + \frac{\mathcal{D}}{2} ,  \label{bottomTs} \\
	z&=1: \hspace*{3 mm}	v_z = 0 \hspace*{3 mm} \mathrm{and}  \hspace*{3 mm} (v_x = v_y = 0  \hspace*{3 mm}\mathrm{or}\hspace*{3 mm} \partial _z v_x = \partial _z v_y = 0),   \label{topTsv} \\
	z&=1: \hspace*{3 mm}  T = -\frac{1}{2} \hspace*{3 mm}\mathrm{or}\hspace*{3 mm} s = -\frac{1}{2 - \mathcal{D}},  \hspace*{3 mm} \mathrm{since}\hspace*{3 mm} T_a = 1 - \frac{\mathcal{D}}{2} . \label{topTs} 
	\end{align}

In astrophysics, the Fourier law for thermal conduction is replaced by a subgrid model of turbulent diffusion for entropy. This has the consequence that the term $\nabla ^2 T$ in equation (\ref{entropya}) is changed for ${\bf \nabla} \cdot \left( \rho _a T_a {\bf \nabla } s   \right)$. In this paper, we will stick to the usual conduction term $\nabla ^2 T$, but the treatment of the 'turbulent diffusivity' would be very similar.

The parameter $\mathcal{D}$ is the one associated with compressibility. Its range, $0 < \mathcal{D} < 2$, covers all cases from the Boussinesq limit ($\mathcal{D} \rightarrow 0$) to the most extreme case of compressibility ($\mathcal{D} \rightarrow 2$) where a temperature of 0~K and a vanishing density are reached at the top of the layer, see equations (\ref{Ta}) and (\ref{rhoa}). 

\section{A minimum principle}
\label{minpple}

In the anelastic liquid approximation (\ref{Ts}), equation (\ref{entropya}) can be re-written using entropy $s$ alone
\begin{equation}
	\rho_a T_a \frac{\mathrm{D}  s}{\mathrm{D} t} = \frac{\mathcal{D}}{Ra } \dot{\varepsilon} : \tau + T_a \nabla^2 s + 2 {\bf \bnabla} T_a \cdot {\bf \bnabla} s , \label{entropya2}
\end{equation}
because $\nabla^2 T_a = 0$, see equation (\ref{Ta}), for our choice of an ideal gas equation of state and uniform gravity. The entropy equation has a suitable form for a maximum principle \citep{picone1929,nirenberg53}. The second order operator is elliptic and even uniformly elliptic as the coefficient $T_a$ is above a positive constant in the whole domain, for any choice of the dissipation parameter $0 < \mathcal{D} < 2$. Furthermore, the term of viscous dissipation $(\mathcal{D}/{Ra })\, \dot{\varepsilon} : \tau$ is positive (or zero) everywhere and at all times. It follows from the maximum principle that $s$ cannot take a value smaller than a value it takes at a boundary or at an initial time. As we are interested in statistically stationary solutions, we argue either that the memory of the initial time is lost or, with more caution, that the initial condition is chosen such that it does not contain values of the entropy $s$ lower than those at the boundaries. We are left with the conclusion that entropy must be larger, everywhere and at all times, than the value assigned at the top boundary 
\begin{equation}
	s \geq - \frac{1}{2-\mathcal{D}}. \label{mins}
\end{equation}

\section{A $\log$-entropy equation}
\label{seclogentropy}

Let us define a constant $s_0 = 1/(2-\mathcal{D}) + 8/(4-\mathcal{D}^2)$, such that $s+s_0$ is always positive from the maximum principle and satisfies
\begin{equation}
        s+s_0 \geq  \frac{8}{4-\mathcal{D}^2}. \label{mins2}
\end{equation}
This choice of $s_0$ will simplify subsequent calculations and will be discussed further in the conclusions. 
Let us divide equation (\ref{entropya2}) by $T_a$ and by $s+s_0$, two positive terms. After re-arranging some terms, we obtain
\begin{equation}
	\rho_a \frac{\mathrm{D}  \mathcal{L} }{\mathrm{D} t} = \frac{\mathcal{D}}{Ra } \frac{ \dot{\varepsilon} : \tau }{T_a (s +s_0 )} + \left| {\bf \bnabla}  \mathcal{L} \right| ^2   + 2 \md _z ( \ln T_a )  \partial _z \mathcal{L} + {\bf \nabla}^2 \mathcal{L} , \label{logentropy}
\end{equation}
where $\mathcal{L} = \ln (s +s_0 )$. Averaging over horizontal directions and in time (denoted by an overline $\overline{X}$ on any variable $X$), and taking into account (\ref{diva}), we obtain an equation for the vertical flux of $\mathcal{L}$
\begin{equation}
	\md _z \Phi _{\mathcal{L}} = \frac{\mathcal{D}}{Ra \ T_a} \overline{\frac{ \dot{\varepsilon} : \tau }{s +s_0 } } + \overline{ \left| {\bf \bnabla}  \mathcal{L} \right| ^2  } + 2  \md _z ( \ln T_a ) \md _z \overline{\mathcal{L}}  , \label{fluxlogentropy}
\end{equation}
where $\Phi _{\mathcal{L}}$ is defined as the average vertical flux of $\mathcal{L}$, at any height $z$, as follows
\begin{equation}
	\Phi _{\mathcal{L}} (z) = -\md _z \overline{\mathcal{L}} + \rho _a \overline{v_z \mathcal{L}}. \label{Philog}
\end{equation}

We shall now average equation (\ref{logentropy}) over the whole layer and in time. This is equivalent to integrating (\ref{fluxlogentropy}) between $z=0$ and $z=1$. Our objective here is to obtain an integral bound on $\left| {\bf \bnabla}  \mathcal{L} \right|^2$. Let us first consider the integral of the last term in (\ref{fluxlogentropy})
\begin{align}
	\int_0^1 2 \md _z ( \ln T_a ) \md _z \overline{\mathcal{L}} \md z &= -2 \mathcal{D} \int_0^1 \frac{\md _z \overline{\mathcal{L}}}{T_a} \md z, \nonumber \\
	&= -2 \mathcal{D} \left( \left[ \frac{\overline{\mathcal{L}}}{T_a}  \right] _0^1  - \int_0^1 \mathcal{D} \frac{\overline{\mathcal{L}}}{T_a^2} \md z    \right) . \label{res}
\end{align}
The first term is known from the boundary conditions (\ref{bottomTs},\ref{topTs}). 
We use the minimum principle (\ref{mins2}), implying $\overline{\mathcal{L}} \geq \ln \left(8/(4-\mathcal{D}^2)\right)$, to bound the last term in (\ref{res}) so that we have
\begin{equation}
	\int_0^1 2 \md _z ( \ln T_a ) \md _z \overline{\mathcal{L}}  \md z \geq \frac{4 \mathcal{D}}{2 + \mathcal{D}} \ln \left( \frac{3}{2} \right) . \label{res2}
\end{equation}
The term $\md _z (\rho _a \overline{v_z \mathcal{L}})$ has no integral contribution because $v_z$ vanishes at the bottom and top, we simply have to evaluate the contribution of the diffusion term $-\md _z \overline{\mathcal{L}}$ in the integral of (\ref{fluxlogentropy}), given that $\partial _z \mathcal{L} = \partial _z (T/T_a )/ (s+s_0 )  $,
\begin{equation} 
	- \left[ \md _z \overline{\mathcal{L}} \right]_0^1 = \frac{Nu}{12} ( 2 + 5 \mathcal{D} ) + \frac{\mathcal{D}}{4} \frac{2 + \mathcal{D} }{2 - \mathcal{D}} +  \frac{\mathcal{D}}{6} \frac{2 - \mathcal{D} }{2 + \mathcal{D}} , \label{diffus}
\end{equation}
where we have denoted $Nu = - \md _z \overline{T} $ the average superadiabatic heat flux injected at the bottom and extracted at the top. The additional heat flux conducted along the adiabat does not affect convection and is here uniform and equal to $\mathcal{D} T_0 / \Delta T_{sa}$ in the same dimensionless scale as the superadiabatic heat flux.

Since the term involving viscous dissipation is positive in (\ref{fluxlogentropy}), combining (\ref{res2}) and (\ref{diffus}) leads to the following bound
\begin{equation}
	\left< \left| {\bf \bnabla}  \mathcal{L} \right| ^2 \right> \leq \frac{Nu}{12} ( 2 + 5 \mathcal{D} ) + \frac{\mathcal{D}}{4} \frac{2 + \mathcal{D} }{2 - \mathcal{D}} +  \frac{\mathcal{D}}{6} \frac{2 - \mathcal{D} }{2 + \mathcal{D}} - \frac{4 \mathcal{D}}{2 + \mathcal{D}} \ln \left( \frac{3}{2} \right) , \label{boundln2}
\end{equation}
where the bracket denotes time and space average (horizontal and vertical), so that $\left< X \right> = \int_0^1 \overline{X} \md z$ for any variable $X$. 
The sum of the last two terms is less than zero for all values of $\mathcal{D}$, hence the sum of the last three terms is less than $\mathcal{D}/(2 - \mathcal{D})$  so that we have
\begin{equation}
        \left< \left| {\bf \bnabla}  \mathcal{L} \right| ^2 \right> \leq \frac{Nu}{12} ( 2 + 5 \mathcal{D} ) + \frac{\mathcal{D} }{2 - \mathcal{D}} . \label{boundln3}
	\end{equation}
This bound is linear in the Nusselt number with a coefficient ranging from $1/6$ at small $\mathcal{D}$ to $1$ when $\mathcal{D}$ reaches its maximum value $2$. In addition, the other term depends on $\mathcal{D}$ only and diverges toward infinity when $\mathcal{D}$ approaches $2$.

We now use the 1D equation (\ref{fluxlogentropy}) to show that the flux of the $\log$-entropy, $\Phi _{\mathcal{L}}$,  
can only increase from the bottom ($z=0$) to a short distance $\delta$ to be defined. 
At $z=0$, the diffusive part of the $\log$-entropy flux, $- \md _z \overline{ \mathcal{L} }$, carries the whole flux 
\begin{equation}
	\Phi _{\mathcal{L}} (0) =  - \md _z \overline{ \mathcal{L} } (0) =  \frac{2 - \mathcal{D}}{6} \left( Nu - \frac{\mathcal{D}}{2 + \mathcal{D}} \right) . \label{philog0}
\end{equation}
We make the assumption that $Nu > \mathcal{D}/(2+\mathcal{D})$, so that the flux (\ref{philog0}) is strictly positive. This assumption is not very restrictive and is satisfied as soon as $Nu > 1/2$. This implies that $\overline{ \mathcal{L}}$ is  locally decreasing at $z=0$. If it kept decreasing at the same rate as in (\ref{philog0}), then the flux $\Phi _{\mathcal{L}}$ would still be carried by diffusion at higher values of $z$. However, $\overline{ \mathcal{L}}$ cannot decrease below the minimum value $\ln \left( 8/(4- \mathcal{D}^2 \right)$, limiting the extension of the diffusive region, and indicating that the convective part of the flux (\ref{Philog}) must take over. We define the height $\delta $ as the smallest value of $z>0$ where the diffusive component, $-\md _z \overline{\mathcal{L} }$, becomes less than half the value $\Phi _{\mathcal{L} } (0)$. The value of $\overline{\mathcal{L} }$ at $z=\delta$ is
\begin{align}
	\overline{\mathcal{L}} (\delta) &= \overline{\mathcal{L}} (0) +  \int_0^\delta \md _z \overline{\mathcal{L}} \,  \md z , \nonumber \\
&\leq \ln \left( \frac{12}{4 - \mathcal{D}^2} \right) - \frac{\delta}{2} \Phi _{\mathcal{L} } (0). \label{deltaf}
\end{align}
The condition that $\overline{ \mathcal{L}}(\delta)$ is above the minimum value of $\mathcal{L}$ -- from (\ref{mins2}) $\mathcal{L} \geq \ln \left( 8 / (4 - \mathcal{D}^2) \right)$ -- implies that $\delta \leq \delta _0$, with
\begin{equation}
	\delta _0 = \frac{12 \ln \left( \frac{3}{2}   \right)}{(2 - \mathcal{D}) \left( Nu - \frac{\mathcal{D}}{2 + \mathcal{D}} \right) }. \label{delta0}
\end{equation}
The definition of $\delta$ implies that $\overline{\mathcal{L}}$ is decreasing everywhere in the range $0 \leq z \leq \delta$. This ensures the positivity of the last term of (\ref{fluxlogentropy}) in that interval, so that $\Phi _{\mathcal{L}} (\delta ) \geq \Phi _{\mathcal{L}} (0)$. Because the diffusive part has been divided by a factor two, 
this means that the convective part of the flux of the $\log$-entropy, at $z=\delta$, must be at least half the boundary flux (\ref{philog0})
\begin{equation}
	\rho _a \overline{v_z \mathcal{L}} (\delta) \geq \frac{2 - \mathcal{D}}{12} \left( Nu - \frac{\mathcal{D}}{2 + \mathcal{D}} \right) . \label{fluxconv}
\end{equation}

Using continuity, the property $\rho _a \leq \rho _{a0}$ and a Cauchy-Schwartz inequality, the convective flux can be bounded as follows
\begin{equation}
	\rho _a \overline{v_z \mathcal{L}} (\delta) \leq \rho _{a0} \overline{v_z (\mathcal{L}-\mathcal{L}_0)} \leq \rho _{a0}  \sqrt{ \overline{v_z^2} } \sqrt{\overline{\left( \mathcal{L} - \mathcal{L}_0 \right) ^2  } } , \label{convflux}
\end{equation}
where $\rho _{a0}$ and $\mathcal{L}_0$ are the density of the adiabatic profile and the value of $\mathcal{L}$ at $z=0$, while $v_z$ and $\mathcal{L}$ are evaluated at $z=\delta $. Now, we use the gradients of $\mathcal{L}$ to bound $\mathcal{L} (\delta )$
\begin{equation}
	\mathcal{L} (\delta) = \mathcal{L}_0 + \int_0^\delta \partial _z \mathcal{L} \md z . \label{boundf}
\end{equation}
Using a Cauchy-Schwartz inequality, we have
\begin{equation}
	\left( \mathcal{L}(\delta) - \mathcal{L}_0 \right) ^2 \leq  \delta  \int_0^\delta \left( \partial _z \mathcal{L} \right) ^2 \md z . \label{boundf2}
\end{equation}
Taking time and horizontal average, extending the last integral over the whole volume and including all gradient components, we obtain
\begin{equation}
	\overline{\left( \mathcal{L}(\delta) - \mathcal{L}_0 \right) ^2} \leq  \delta  \left< \left| \bnabla \mathcal{L} \right| ^2 \right> \leq  \delta _0 \left< \left| \bnabla \mathcal{L} \right| ^2 \right> , \label{boundf3}
\end{equation}
where $\left< \left| \bnabla \mathcal{L} \right| ^2 \right>$ is itself bounded from (\ref{boundln3}). 
Following the same steps, we can bound $v_z$, at any height $y$, as 
\begin{equation}
	\overline{ v_z^2 } (y)  \leq y  \int _0^y \overline{(\partial _z v_z )^2} \md z . \label{boundvz}
\end{equation}
Now, we need to relate $ \overline{(\partial _z v_z )^2} $ to the mean viscous dissipation. From the general expression of viscous dissipation with zero bulk viscosity \citep{Landau} 
\begin{equation}
	\dot{\varepsilon} : \tau = \frac{1}{2} \sum_{i=1}^3 \sum_{j=1}^3 \left[ \partial _i v_j + \partial _j v_i - \frac{2}{3} (\partial _k v_k ) \delta _{ij}   \right]^2, \label{generaldiss}
\end{equation}
retaining only the three 'diagonal' terms $i=j$ among the nine terms, finally dropping $2(\partial _x v_x )^2$ and $2(\partial _y v_y )^2$, we derive
\begin{equation}
	\dot{\varepsilon} : \tau  \geq 2 (\partial _z v_z )^2 - \frac{2}{3} (\partial _k v_k )^2. \label{dissv2}
	\end{equation}
Using the anelastic equation of continuity (\ref{diva}), this leads to an upper bound of $(\partial _z v_z )^2$
\begin{equation}
	(\partial _z v_z )^2 \leq \frac{1}{2} \dot{\varepsilon} : \tau  + \frac{1}{3} \frac{\mathcal{D}^2 v_z^2}{(\gamma -1 )^2 T_a^2} ,  \label{dissv3}
	\end{equation}
which is substituted in (\ref{boundvz}) to obtain
\begin{equation}
	\overline{v_z^2} (y) \leq \frac{y}{2} \left< \dot{\varepsilon} : \tau \right> + \frac{\mathcal{D}^2}{3 (\gamma -1)^2}  y \int _0^y \frac{\overline{v_z^2}}{T_a^2}  \md z . \label{vz2}
	        \end{equation}
We now use this equation for $y \leq \delta \leq \delta _0$, assuming that $\delta _0 \leq 1/2$ so that $1/T_a^2$ in the integral above is smaller than $1$. Integrating from $0$ to $\delta$ leads to the bound
\begin{equation}
	\int _0^\delta \overline{v_z^2} (y) \md y \leq \frac{\delta ^2}{4} \left< \dot{\varepsilon} : \tau \right> + \frac{\mathcal{D}^2}{3 (\gamma -1)^2} \frac{\delta ^2}{2} \int _0^\delta \overline{v_z^2} (z) \md z, \label{vz2bis}
\end{equation}
which is used to express $\int _0^\delta \overline{v_z^2} (z) \md z$ in terms of $\left< \dot{\varepsilon} : \tau \right>$ so that (\ref{vz2}) can finally be written at $z=\delta$  
\begin{equation}
	\overline{v_z^2} (\delta ) \leq \frac{\delta _0 }{2} \left< \dot{\varepsilon} : \tau \right> \frac{1}{1 - \frac{\mathcal{D}^2 \delta _0^2}{6 (\gamma -1)^2 }} . \label{vz2ter}
	\end{equation}

With (\ref{boundln3}), (\ref{delta0}), (\ref{convflux}), (\ref{boundf3}), (\ref{vz2ter}), we can essentially write (\ref{fluxconv}) as a lower bound for the viscous dissipation
\begin{equation}
	\left< \dot{\varepsilon} : \tau \right> \geq \frac{2 \, Nu^3 \, (2 - \mathcal{D})^4 }{12^3 \left( \ln \frac{3}{2} \right)^2 \rho _{a0}^2 (2 + 5 \mathcal{D}) } \frac{\left[ 1 - \frac{\mathcal{D}}{Nu \, (2 + \mathcal{D})}  \right]^4 \left[ 1 - \frac{24 \left( \ln \frac{3}{2} \right)^2 \mathcal{D}^2}{(2-\mathcal{D})^2 \left( Nu - \frac{\mathcal{D}}{2 + \mathcal{D}} \right)^2 (\gamma -1 )^2 } \right]}{1 + \frac{12 \mathcal{D}}{Nu (2 - \mathcal{D} ) (2 + 5 \mathcal{D})}} . \label{lowdiss} 
\end{equation}
This bound is valid as long as the thickness $\delta _0$ defined in equation  (\ref{delta0}) is less than $1/2$. 
For very large Nusselt numbers (more precisely $(2- \mathcal{D}) Nu \gg 1$), this lower bound becomes 
\begin{equation}
	\left< \dot{\varepsilon} : \tau \right> \geq \frac{2 \, Nu^3 \, (2 - \mathcal{D})^4 }{12^3 \left( \ln \frac{3}{2} \right)^2 \rho _{a0}^2 (2 + 5 \mathcal{D}) }. \label{approxdiss}
\end{equation}


\section{Upper bound on viscous dissipation for a given heat flux}
\label{viscousdiss}

We have thus obtained (\ref{lowdiss}) a lower bound of the viscous dissipation $\left< \dot{\varepsilon} : \tau \right>$ for a given heat flux $Nu$. Now, considering the classical entropy budget, we are going to obtain an upper bound for $\left< \dot{\varepsilon} : \tau \right>$. 
Let us divide the governing equation (\ref{entropya}) by $T_a$, rearrange the diffusion term, and obtain the usual anelastic entropy equation 
\begin{equation}
	\rho_a \frac{\mathrm{D}  s}{\mathrm{D} t} = \frac{\mathcal{D}}{Ra } \frac{\dot{\varepsilon} : \tau }{T_a} + \frac{\bnabla T \cdot \bnabla T_a}{T_a^2} + \bnabla \cdot \left( \frac{ \bnabla T }{T_a } \right), \label{entropyabis}
\end{equation}
The time and space average of the second term on the right-hand side can be evaluated as follows
\begin{align}
	\left< \frac{\bnabla T \cdot \bnabla T_a}{T_a^2} \right> &= - \int _0^1 \frac{\mathcal{D}}{T_a^2} \md _z \overline{T} \md z = \left[ - \frac{\mathcal{D}}{T_a^2} \overline{T}   \right] _0^1 + \int_0^1 \md _z \left( \frac{\mathcal{D}}{T_a^2}  \right) \overline{T} \md z \nonumber \\
	&= \frac{4 \mathcal{D} (4 + \mathcal{D}^2)}{(4 - \mathcal{D}^2)^2} + 2 \mathcal{D}^2 \int_0^1 \frac{1}{T_a^2} \frac{\overline{T} }{T_a} \md z  .     \label{gradTagraT}
\end{align}
Once again, the maximum principle applying to the entropy variable $s=T/T_a$ is used to bound the second term. 
\begin{align}
	\left< \frac{\bnabla T \cdot \bnabla T_a}{T_a^2} \right> & \geq \frac{4 \mathcal{D} (4 + \mathcal{D}^2)}{(4 - \mathcal{D}^2)^2} - \frac{2 \mathcal{D}^2}{2 - \mathcal{D}} \int_0^1 \frac{1}{T_a^2} \md z \nonumber \\
	& \geq \frac{4 \mathcal{D} \left( 4 - 4 \mathcal{D} - \mathcal{D}^2 \right) }{(4 - \mathcal{D}^2 )^2 } . \label{gradTagraTbis}
\end{align}
With that bound (\ref{gradTagraTbis}), integrating equation (\ref{entropyabis}) in space and time leads to an upper bound 
\begin{equation}
	\frac{\mathcal{D}}{Ra } \left< \frac{\dot{\varepsilon} : \tau }{T_a} \right> \leq \frac{4 \mathcal{D} }{4 - \mathcal{D}^2 } Nu - \frac{4 \mathcal{D} \left( 4 - 4 \mathcal{D} - \mathcal{D}^2 \right) }{(4 - \mathcal{D}^2 )^2 } . \label{disssurTa}
\end{equation}
As $T_a \leq 1 + \mathcal{D}/2$, this becomes an upper bound on viscous dissipation
\begin{equation}
	\left< \dot{\varepsilon} : \tau \right> \leq \frac{2 \, Ra }{2 - \mathcal{D} } Nu \left[ 1 - \frac{ 4 - 4 \mathcal{D} - \mathcal{D}^2 }{(4 - \mathcal{D}^2 ) Nu } \right] . \label{dissUB}
\end{equation}
In the limit of large Nusselt numbers (more precisely $(2- \mathcal{D}) Nu \gg 1$), this upper bound becomes
\begin{equation}
\left< \dot{\varepsilon} : \tau \right> \leq \frac{2 \, Ra }{2 - \mathcal{D} } Nu  . \label{dissUBapprox}
\end{equation}


\section{Obtaining an upper bound on the heat flux}
\label{upperbound}

Combining the upper bound (\ref{dissUB}) and the lower bound (\ref{lowdiss}) leads to the following inequality
\begin{equation}
	Ra \geq \frac{Nu^2 \, (2 - \mathcal{D})^5 }{12^3 \left( \ln \frac{3}{2} \right)^2 \rho _{a0}^2 (2 + 5 \mathcal{D}) } \frac{\left[ 1 - \frac{\mathcal{D}}{Nu \, (2 + \mathcal{D})}  \right]^4 \left[ 1 - \frac{12^2 \left( \ln \frac{3}{2} \right)^2 \mathcal{D}^2}{6(2-\mathcal{D})^2 \left( Nu - \frac{\mathcal{D}}{2 + \mathcal{D}} \right)^2 (\gamma -1 )^2 } \right]}{\left[1 + \frac{12 \mathcal{D}}{Nu (2 - \mathcal{D} ) (2 + 5 \mathcal{D})}\right] \left[ 1 - \frac{ 4 - 4 \mathcal{D} - \mathcal{D}^2 }{(4 - \mathcal{D}^2 ) Nu }  \right] } . \label{NuRaD} 
\end{equation}
This bound is valid when the thickness $\delta _0$ is less than $1/2$. In the limit $(2- \mathcal{D}) Nu \gg 1$, we just have
\begin{equation}
	Ra \geq \frac{Nu^2 \, (2 - \mathcal{D})^5 }{12^3 \left( \ln \frac{3}{2} \right)^2 \rho _{a0}^2 (2 + 5 \mathcal{D}) } , \label{NuRaDapprox}
\end{equation}
which may be re-written 
\begin{equation}
	Nu \leq 12^{3/2} \left( \ln \frac{3}{2} \right) \rho _{a0} (2 + 5 \mathcal{D})^{\frac{1}{2}}  \frac{Ra^{\frac{1}{2}}}{(2 - \mathcal{D})^{\frac{5}{2}}} . \label{NuRaDapprox2}
\end{equation}
When $\mathcal{D}$ varies from $0$ to $2$, the value of $\rho _{a0}$ varies from $1$ to $2.5$ (perfect monatomic gas) and $2+5\mathcal{D}$ is less than $12$, so that we have a simpler bound of the form
\begin{equation}
        Nu < 146 \, \frac{Ra^{\frac{1}{2}}}{(2 - \mathcal{D})^{\frac{5}{2}}} . \label{NuRaDapprox3}
\end{equation}

\begin{figure}
	\begin{center}
		\includegraphics[width=12 cm, keepaspectratio]{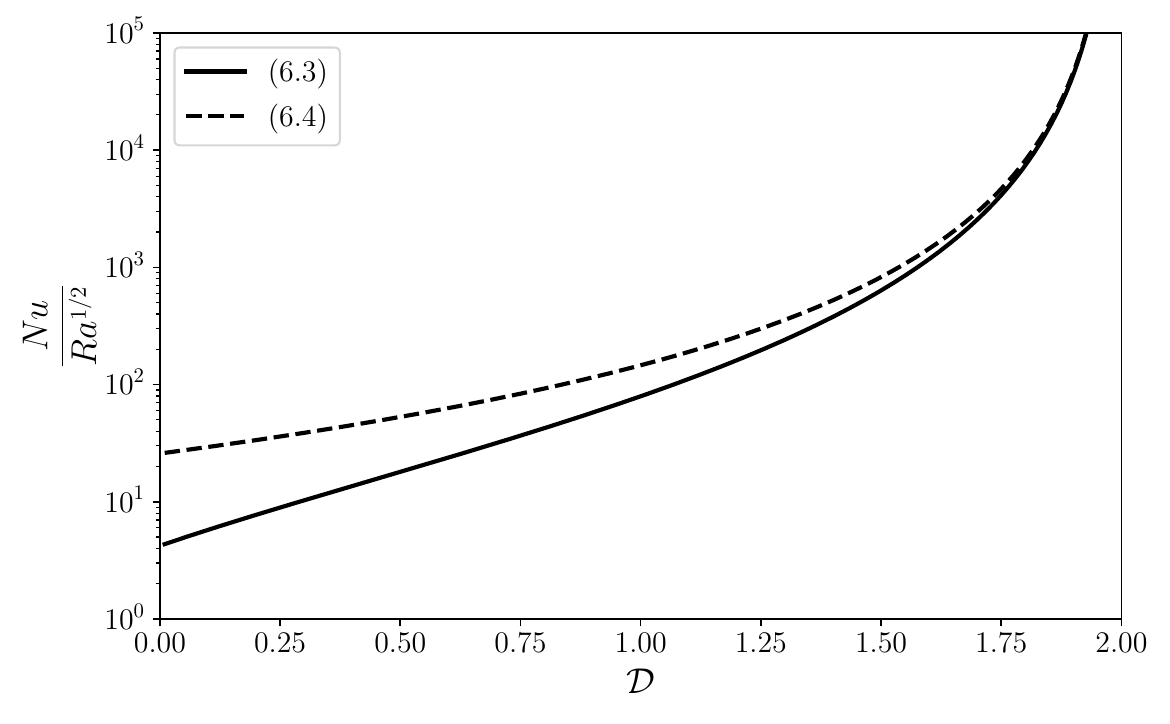}
		\caption{Prefactor of $Ra^{1/2}$ in the Nusselt bound, in the limit of large values of $(2-\mathcal{D}) Nu$ for a perfect gas with $\gamma =5/3$, computed from (\ref{NuRaDapprox2}). The simpler expression computed from (\ref{NuRaDapprox3}) is also shown as a function of $\mathcal{D}$.  }
		\label{figure}
	\end{center}
\end{figure}

We plot the $Ra^{1/2}$ prefactors of these Nusselt laws in Fig.~\ref{figure}, using a logarithmic scale since the singularity at $\mathcal{D}=2$ leads to large values of $Nu$. 

\section{Conclusions}

We have obtained an upper bound for the heat transfer in a compressible model of convection known as the anelastic liquid approximation. The bound is expressed in an algebraic form (\ref{NuRaD}), valid under the condition that the distance $\delta _0$ defined in (\ref{delta0}) is less than $1/2$, which is easily reached at moderate Nusselt numbers. It takes the simpler expression (\ref{NuRaDapprox2}) in the limit of large Nusselt numbers. This simpler expression can itself be bounded by $Nu < 146\, Ra^{1/2} / (2-\mathcal{D})^{5/2}$. 
The same method, with nearly the same result, applies to the anelastic model used in astrophysics, where thermal conduction is modelled using the gradient of entropy. 

In order to obtain this bound, we have introduced an unusual quantity, the logarithm of entropy (shifted with a constant so that it is positive everywhere), and have shown that it obeys a equation similar to that of entropy. Its flux has a conduction term and a convective term and its sources are positive (or at least bounded from below). The difference with entropy is that it is possible to derive an $L^2$ upper bound for the gradients of this quantity while we could not do so for the gradients of entropy. 

Importantly, obtaining a bound relies heavily on the existence of a maximum principle for entropy (actually, a minimum). Although entropy is bounded from below only, this limit enables us to bound the diffusive part of the flux close to the hot boundary and also to bound the sources of entropy (or its logarithm) from below. 

There is a degree of freedom in the choice of the constant $s_0$ added to entropy in order to make it strictly positive. We have chosen it so that the minimum value of $s+s_0$ is $8/(4-\mathcal{D}^2)$, see equation (\ref{mins2}). That choice affects the final upper bound of Nusselt number and we can choose $s_0$, for each value of $\mathcal{D}$, to obtain the lowest possible upper bound. Although our choice is not exactly the optimal choice, one cannot improve the final upper bound of Nusselt number by more than 25~\% with another choice for $0.1 < \mathcal{D} < 2$, while a better choice could lower the upper bound by a factor 2 near $\mathcal{D} = 0$. Our value has the advantage of making the algebra simpler. 

There is another degree of freedom concerning the fraction of the conduction term needed to define the thickness $\delta _0$. We have decided to consider when $1/2$ of the flux must be carried by convection, but we could have taken any fraction between $0$ and $1$. Actually, this choice of $1/2$ leads to the best final bound in our case. In the Boussinesq case, \cite{seis2015} found that a fraction $1/3$ was the optimum, but both the governing equations and the nature of the flux are different (heat flux versus flux of $\log$-entropy). 

The bound (\ref{NuRaD}) and its approximations at large Nusselt numbers (\ref{NuRaDapprox2})  or (\ref{NuRaDapprox3}) are not expected to be very tight. In the limit $\mathcal{D} \rightarrow 0$, where the anelastic model should converge toward the Boussinesq model, we can readily see that it is less tight than the original bound by Howard by a factor nearly $20$ (see Fig.~\ref{figure}). Concerning large values of $\mathcal{D}$, our bound is made very large owing to the divergent factor $(2-\mathcal{D})^{-5/2}$. We think this is due to our inability to track the $\log$-entropy flux near the cold boundary and, more fundamentally, this originates from the lack of an upper bound for entropy (we only have a lower bound). However, such a trend is not observed in numerical calculations, on the contrary an increase of the dissipation number seems to lead to a decrease of the heat flux \citep{cdadlr2019}. 

Among the possible extensions of this work to other models, it would be interesting to consider fluids of infinite Prandtl number. In that case and in the Boussinesq model, tight upper bounds on the heat flow have been obtained \citep{doering_otto_reznikoff_2006}. Also in the Boussinesq limit and when Coriolis forces are taken into account, upper bounds have been derived \citep{PhysRevFluids.7.093501,tilgner_2022}. Extending these results to a compressible model of convection would be relevant to planetary convection. For that purpose, we need to consider different models of equation of state for condensed matter, idealized \citep{acdlr2022} or more realistic \citep{ralcd2022,ra2022} concerning compressible convection in planetary interiors. \\

\noindent
{\bf Declaration of Interests}. The authors report no conflict of interest.\\

\bibliography{local}

\begin{thebibliography}{17}
\providecommand{\natexlab}[1]{#1}
\providecommand{\url}[1]{\texttt{#1}}
\expandafter\ifx\csname urlstyle\endcsname\relax
  \providecommand{\doi}[1]{doi: #1}\else
  \providecommand{\doi}{doi: \begingroup \urlstyle{rm}\Url}\fi

\bibitem[Alboussière et~al.(2022)Alboussière, Curbelo, Dubuffet, Labrosse,
  and Ricard]{acdlr2022}
T.~Alboussière, J.~Curbelo, F.~Dubuffet, S.~Labrosse, and Y.~Ricard.
\newblock A playground for compressible natural convection with a nearly
  uniform density.
\newblock \emph{Journal of Fluid Mechanics}, 940:\penalty0 A9, 2022.

\bibitem[Anufriev et~al.(2005)Anufriev, Jones, and Soward]{ajs2005}
A.P. Anufriev, C.A. Jones, and A.M. Soward.
\newblock {The {B}oussinesq and anelastic liquid approximations for convection
  in the {E}arth’s core}.
\newblock \emph{Phys. {E}arth and {P}lanet. {I}nt.}, 152:\penalty0 163--190,
  2005.

\bibitem[Curbelo et~al.(2019)Curbelo, Duarte, Alboussi\`ere, Dubuffet,
  Labrosse, and Ricard]{cdadlr2019}
J.~Curbelo, L.~Duarte, T.~Alboussi\`ere, F.~Dubuffet, S.~Labrosse, and
  Y.~Ricard.
\newblock {N}umerical solutions of compressible convection with an infinite
  {P}randtl number: comparison of the anelastic and anelastic liquid models
  with the exact equations.
\newblock \emph{Journal of Fluid Mechanics}, 873:\penalty0 646--687, 2019.

\bibitem[Doering and Constantin(1996)]{DoeringConstantin96}
C.R. Doering and P.~Constantin.
\newblock Variational bounds on energy dissipation in incompressible flows.
  {III}. {C}onvection.
\newblock \emph{Phys.\ Rev.\,E}, 53\penalty0 (6):\penalty0 5957--5981, 1996.

\bibitem[Doering et~al.(2006)Doering, Otto, and
  Reznikoff]{doering_otto_reznikoff_2006}
C.R. Doering, F.~Otto, and M.G. Reznikoff.
\newblock Bounds on vertical heat transport for infinite-{P}randtl-number
  {R}ayleigh–{B}\'enard convection.
\newblock \emph{Journal of Fluid Mechanics}, 560:\penalty0 229--241, 2006.

\bibitem[Emanuel(1998)]{EMANUEL19981313}
G.~Emanuel.
\newblock Bulk viscosity in the navier–stokes equations.
\newblock \emph{International Journal of Engineering Science}, 36\penalty0
  (11):\penalty0 1313--1323, 1998.

\bibitem[Howard(1963)]{howard63}
L.N. Howard.
\newblock Heat transport by turbulent convection.
\newblock \emph{Journal of Fluid Mechanics}, 17\penalty0 (3):\penalty0
  405--432, 1963.

\bibitem[Joseph(1976)]{joseph76}
D.~Joseph.
\newblock \emph{Stability of Fluid Motions}.
\newblock Springer-Verlag, 1976.

\bibitem[Landau and Lifshitz(1966)]{Landau}
L.~Landau and E.~Lifshitz.
\newblock \emph{M\'ecanique des fluides}.
\newblock MIR Moscou, 1966.

\bibitem[Lantz and Fan(1999)]{lf1999}
S.R. Lantz and Y.~Fan.
\newblock Anelastic magnetohydrodynamic equations for modeling solar and
  stellar convection zones.
\newblock \emph{Astrophys. Journal}, 121:\penalty0 247--264, 1999.

\bibitem[Nirenberg(1953)]{nirenberg53}
L.~Nirenberg.
\newblock A strong maximum principle for parabolic equations.
\newblock \emph{Comm. Pure Appl. Math.}, 6:\penalty0 167--177, 1953.

\bibitem[Picone(1929)]{picone1929}
M.~Picone.
\newblock Maggiorazione degli integrali delle equazioni totalmente paraboliche
  alle derivate parziali del secondo ordine.
\newblock \emph{Ann. Mat. Pura Appl.}, 7:\penalty0 145--192, 1929.

\bibitem[Ricard and Alboussi\`ere(2023)]{ra2022}
Y.~Ricard and T.~Alboussi\`ere.
\newblock Compressible convection in super-earths.
\newblock \emph{Physics of the Earth and Planetary Interiors}, 341:\penalty0
  107062, 2023.
\newblock ISSN 0031-9201.

\bibitem[Ricard et~al.(2022)Ricard, Alboussière, Labrosse, Curbelo, and
  Dubuffet]{ralcd2022}
Y.~Ricard, T.~Alboussière, S.~Labrosse, J.~Curbelo, and F.~Dubuffet.
\newblock {Fully compressible convection for planetary mantles}.
\newblock \emph{Geophysical Journal International}, 230\penalty0 (2):\penalty0
  932--956, 03 2022.
\newblock ISSN 0956-540X.

\bibitem[Seis(2015)]{seis2015}
C.~Seis.
\newblock Scaling bounds on dissipation in turbulent flows.
\newblock \emph{Journal of Fluid Mechanics}, 777:\penalty0 591--603, 2015.

\bibitem[Tilgner(2022{\natexlab{a}})]{PhysRevFluids.7.093501}
A.~Tilgner.
\newblock Bounds for rotating convection at infinite {P}randtl number from
  semidefinite programs.
\newblock \emph{Phys. Rev. Fluids}, 7:\penalty0 093501, Sep 2022{\natexlab{a}}.

\bibitem[Tilgner(2022{\natexlab{b}})]{tilgner_2022}
A.~Tilgner.
\newblock Bounds for rotating {R}ayleigh–{B}énard convection at large
  {P}randtl number.
\newblock \emph{Journal of Fluid Mechanics}, 930:\penalty0 A33,
  2022{\natexlab{b}}.

\end{thebibliography}

\end{document}